\documentclass[aps,prl,twocolumn,showpacs]{revtex4-1}

\usepackage{graphicx}
\usepackage{epsfig}

\begin{document}

\input{epsf}

\title{Formation of Rydberg macrodimers and their properties}
\author{Nolan Samboy$^1$}
\author{Jovica Stanojevic$^2$}
\author{Robin C\^{o}t\'{e}$^1$}

\affiliation{$^1$Physics Department, University of Connecticut,
             2152 Hillside Rd., Storrs, CT 06269-3046}
\affiliation{$^2$ Max Planck Institute for the Physics of Complex 
             Systems, N\"othnitzer Strasse 38, D-01187 Dresden, Germany}

\date{\today}
\begin{abstract}

We investigate the interaction between two rubidium atoms in highly excited 
Rydberg states, and find that very long-range potential wells exist. These 
wells are shown to support many bound states. We calculate the properties of
the wells and bound levels, and show that their lifetimes are limited 
by that of the constituent Rydberg atoms. We also show how these 
$\mu$m-size bound states can be populated {\it via} photoassociation (PA), and how
the signature of the ad-mixing of various $\ell$-character producing the potential 
wells could be probed. We discuss how sharp variations of the PA rate 
could act as a switching mechanism with potential application 
to quantum information processing.
\end{abstract}


\maketitle

In recent years, the strong interactions between Rydberg atoms, due
to their exaggerated properties \cite{Gallagher}, have been detected 
experimentally \cite{Anderson, Mourachko}, and led to proposals for 
quantum computing \cite{Saffman-RMP}, {\it e.g.} to achieve fast 
quantum gates \cite{jaksch00,grangier02} or to study 
quantum random walks \cite{cote-qrw}. One effect of these interactions is the 
excitation blockade \cite{lukin01}, where a Rydberg atom prevents the excitation
of nearby atoms \cite{tong04,singer04,Liebisch,vogt06,Heidemann08}. 
Recently, dipole blockade between two atoms has been observed in microtraps 
\cite{grangier08,saffman09}, and a C-NOT gate implemented \cite{Saffman10}.
Another signature of these strong interactions is the molecular resonance 
features in excitation spectra, first observed in Rb \cite{farooqi03} and in
Cs atoms \cite{overstreet07}. Other molecular features involving Rydberg 
excitations, such as the so called {\it trilobite} and {\it butterfly} 
states, where one atom is in its ground state and another in a Rydberg 
state \cite{trilobites}, have recently 
been detected \cite{pfau}, while polyatomic molecules involving Rydberg 
electrons have also been studied \cite{Rost,Sadeghpour}.

In this article, we investigate the existence of long-range potential wells
produced by two atoms excited into Rydberg states.
As opposed to previous predictions of {\it macrodimers} \cite{macro-old} --
doubly-excited Rydberg molecules -- with very shallow wells due to 
induced van der Waals (vdW) interactions, the
macrodimers we study here are due to the strong
mixing of $\ell$-characters of various Rydberg states. As we will see, these
molecular wells are deeper (a few GHz), but still very extended with an
equilibrium separation of 1 $\mu$m or so. We concentrate on wells
existing in the vicinity of two rubidium (Rb) atoms excited to the $np$ Rydberg 
state, for which the relevant molecular symmetries are
$0_g^+$, $0_u^-$, and $1_u$ \cite{Jovica}. 


\begin{figure}[h]
  \centerline{\epsfxsize=3.25in\epsfclipon\epsfbox{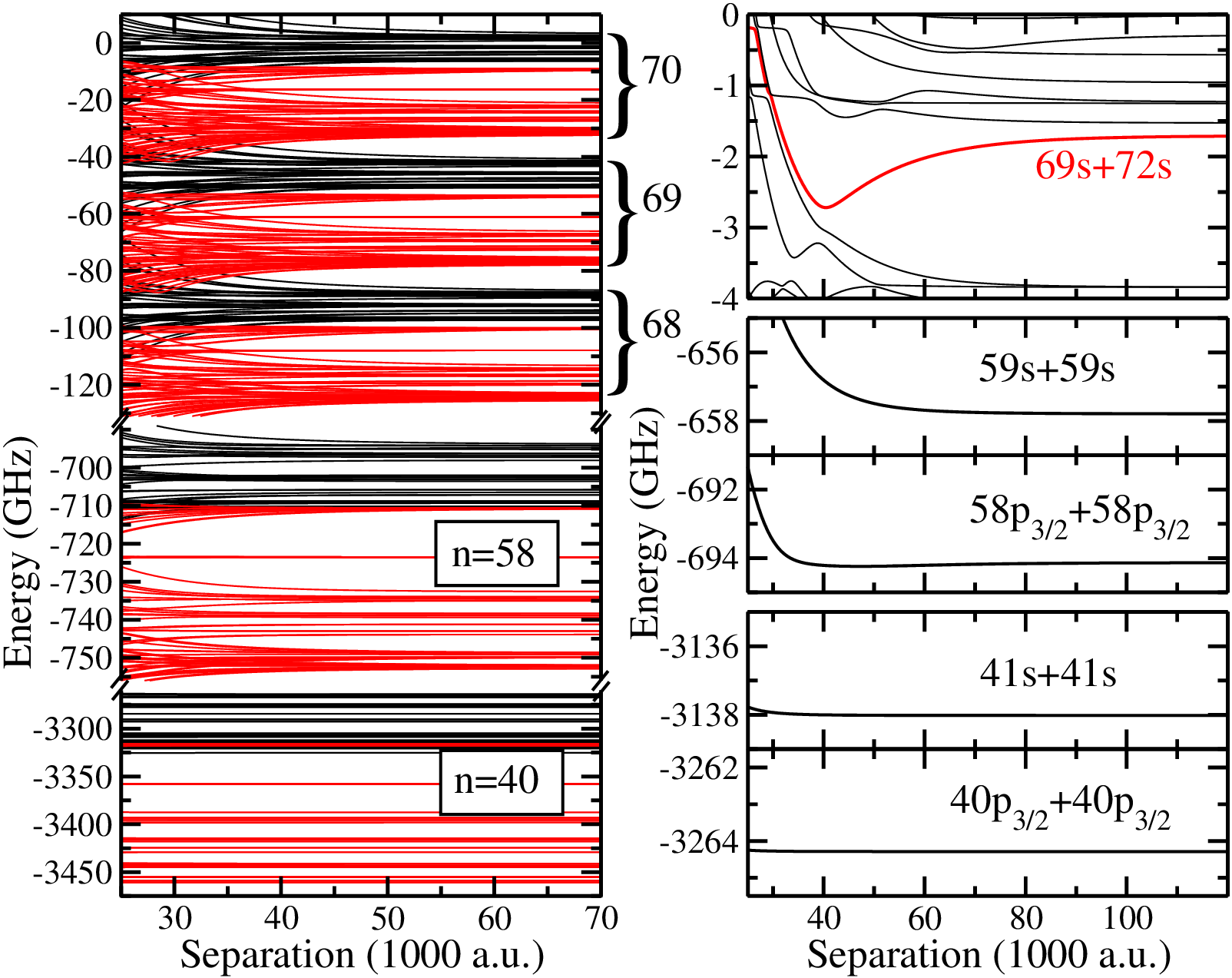}}
\caption{(Color online) $0^{+}_g$ molecular curves. Left panel:
         curves correlated to $n_1\ell_1 +n_2\ell_2$ for $n\sim 70, 69$, 
         and 68 (top plot), 58 (middle) and 40 (bottom); black curves
         correspond to diagonalization near $np+np$ asymptotes, and 
         red curves near $ns+ns$. Right panel: enlargements 
         near the $70p_{3/2}+70p_{3/2}$ asymptote (top: with a well correlated 
         to $69s+72s$), and the $R$ dependence of lower asymptotes (middle: 
         $n\sim 58$; bottom: $n \sim 40$). Those last plots illustrate
         how the potential curves become flat as $n$ decreases.}
	 \label{EF00}
\end{figure}

We build our long-range molecular potential curves following the procedure 
described in \cite{Jovica}; we consider two free Rydberg atoms 
in states $|a\rangle\equiv|n,\ell,j,m_j\rangle$ and 
$|a'\rangle\equiv|n',\ell',j',m'_j\rangle$, where
$n$ is the principal quantum number, $\ell$ the orbital angular momentum, and $m_j$ 
is the projection of the total angular momentum $\vec{j}=\vec{\ell}+\vec{s}$ 
onto a quantization axis (chosen in the $z$-direction for convenience).
The long-range molecular potential curves in Hund's case (c) are calculated 
directly by diagonalizing the interaction Hamiltonian consisting of both 
the Rydberg-Rydberg interaction and the atomic fine structure \cite{Jovica}. 
The basis states are built as follows:
\begin{eqnarray}
|a,a';\Omega_{g/u}\rangle 
\sim |a\rangle_1 |a'\rangle_2 
- p(-1)^{\ell+\ell'}|a'\rangle_1 |a\rangle_2
\label{basis}
\end{eqnarray}
where $\Omega = m_j + m'_j$ is the projection of the total angular momentum 
on the molecular axis and is conserved. The quantum number $p$, describing the
symmetry property under inversion, is 1(-1) for $g(u)$ states. For $\Omega=0$, 
the states' symmetry under reflection must also be considered \cite{Jovica}.
The basis (\ref{basis}) assumes no overlap of charge distributions between the 
two atoms, so the long-range interaction can be expanded as an 
inverse power series of the nuclear separation distance $R$ \cite{Jovica}.

Fig.~\ref{EF00} shows the results of the diagonalized interaction matrix for 
the $0_g^{+}$ symmetry near various $n\ell + n\ell$ asymptotes.  The left panel
shows curves for several states located around $np_{3/2}+np_{3/2}$ (in black)
and $ns+ns$ (in red) near $n=70$, 69, 68, and lower asymptotes as well 
($n\sim 58$ and $40$). These plots depict the strong mixing of 
different $\ell$-characters due mainly to the dipole-dipole interactions 
(although higher orders are also present). The right panel enlarges
a major feature arising from the $\ell$-mixing near $70p_{3/2}+70p_{3/2}$, 
namely a large potential well (between 0 and -2.5 GHz) with a depth 
$D_e\sim 1$ GHz measured from its minimum (equilibrium separation) 
$R_e$ at roughly 40,500 $a_0$ ($a_0$: Bohr radius), {\it i.e.} over 
2 $\mu$m. The $R$-dependence for lower 
asymptotes is also shown on the same panel; as $n$ decreases
the potential curves become flatter (for same $R$ interval and energy scale).
This is particularly true as $n$ reaches 40 or so.

\begin{figure}[h]
  \centerline{\epsfxsize=3.25in\epsfclipon\epsfbox{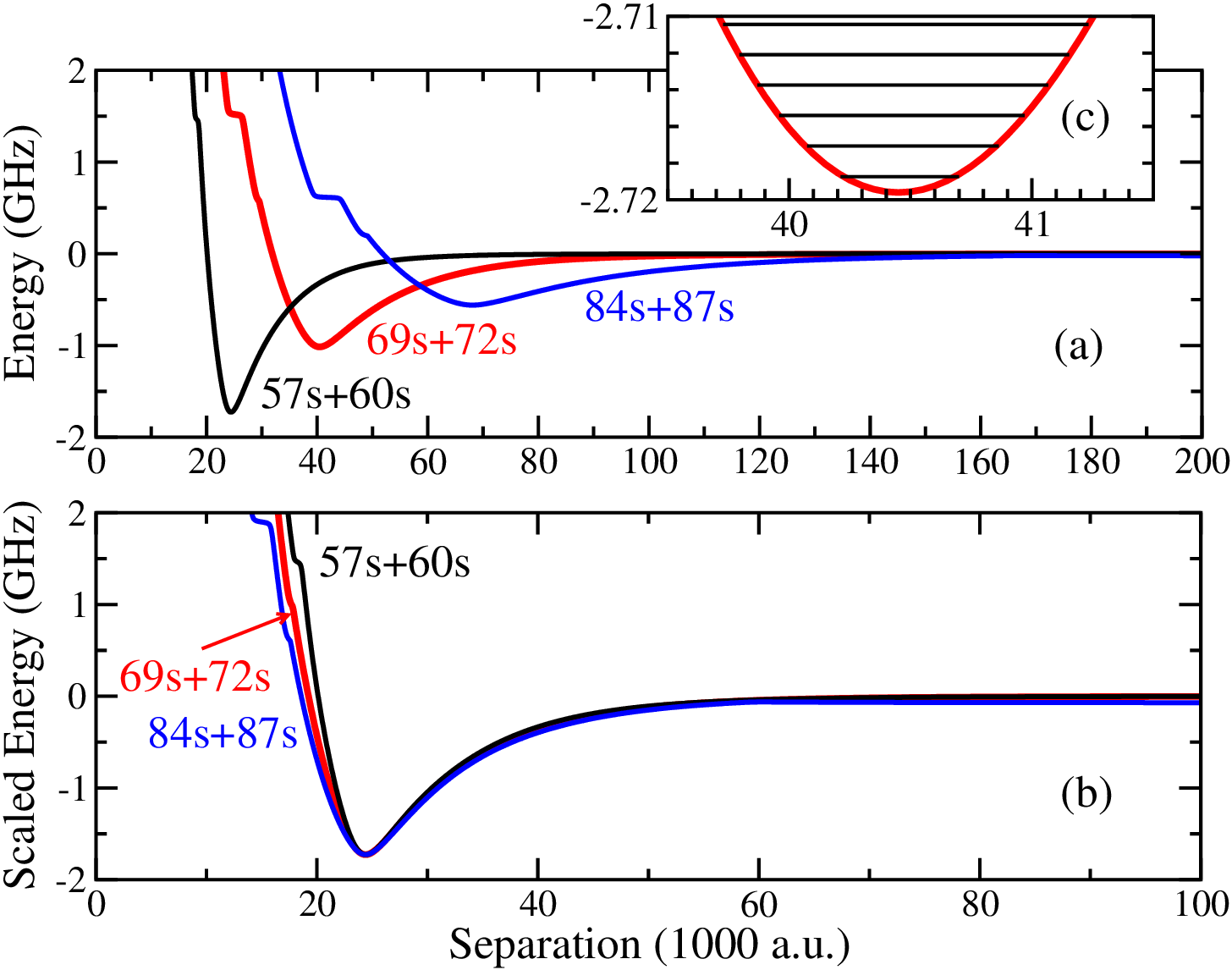}}
\caption{(Color online) $0^{+}_g$ wells correlated to the $(n-1)s+(n+2)s$ 
            asymptote for $n=58$, 70 and 85: (a) un-scaled, and (b) scaled to $n=58$
            with $D_e\sim n^{-3}$ and $R_e\sim n^{2.5}$. In (c), we show the
            deepest bound levels for $n=70$ (original scale).}
	   \label{fig:scaling}
\end{figure}

The long-range well correlated to the $(n-1)s+(n+2)s$ asymptote is a general
feature for this system; we depict a few cases in Fig.~\ref{fig:scaling}(a). 
The well depth $D_e$ and equilibrium separation $R_e$ scale 
roughly as $n^{-3}$ and $n^{2.5}$ (Fig.~\ref{fig:scaling}(b)), respectively, in
good agreement with the $D_e\!\propto\! n^{-3}$ and $R_e\!\propto\!  n^{7/3}$ 
scalings  expected for a dominantly dipolar coupling between states. 
This last result can be obtained by keeping only the major $n$-dependence, 
so that the interaction matrix between two asymptotes has 
the form $n^{-3} {\hat h}_0 \!+\! n^{4} R^{-3}{\hat h}_1 \!$ where 
${\hat h}_0$ and ${\hat h}_1$ contain the energy spacings between the 
asymptotes and the strengths of the dipole-dipole coupling, respectively. 
If we define $\rho\!\equiv\!n^{7} R^{-3}$ and rewrite the interaction matrix
as $n^{-3} ({\hat h}_0  \!+\! \rho {\hat h}_1)$, it becomes clear that its 
eigenvalues $E_i$ take the form $E_i\!=\! n^{-3} f_i(\rho)$. Since 
$\partial E_i/\partial R \!=\!-3 n^{4}  R^{-4}f'_i(\rho)\!=\!0$ at an extremum,
we must have $f'_i(\rho)\!=\!0$, which is only fulfilled for some particular 
values $\rho_e$. In the parameter space of $R$ and $n$, this leads to
$\rho_e\!=\!n^{7} R^{-3}|_{R_e}\!=\!{\rm const.}$ so that $R_e\!\propto\! n^{7/3}$,
and $f_i(\rho_e)\!=\!n^{3}E_i|_{D_e}\!=\!{\rm const.}$ so that $D_e\!\propto\!  n^{-3}$.
The difference between the analytical and numerical scalings reflects the more 
complex nature of the coupling than the assumed dipolar interaction.


We return to the curve correlated to the $69s+72s$ asymptote in the $0_g^+$ 
symmetry. Its well is deep enough to support several bound molecular levels, 
a few of which are listed in Table ~\ref{Lifetimes} and shown in Fig.~\ref{fig:scaling}(c).  
Since the levels are separated 
by a few MHz, they could be detected by spectroscopic means. 
Oscillation periods of a few $\mu$s, related to the MHz frequency of these
deepest levels, are rapid enough to allow several oscillations during the lifetime of
the Rydberg atoms (roughly a few 100 $\mu$s for $n\sim 70$). The classical inner
and outer turning points of the levels indicate very extended molecules, hence
we keep the term {\it macrodimer} to describe them.

\begin{figure}[h]
  \centerline{\epsfxsize=3.25in\epsfclipon\epsfbox{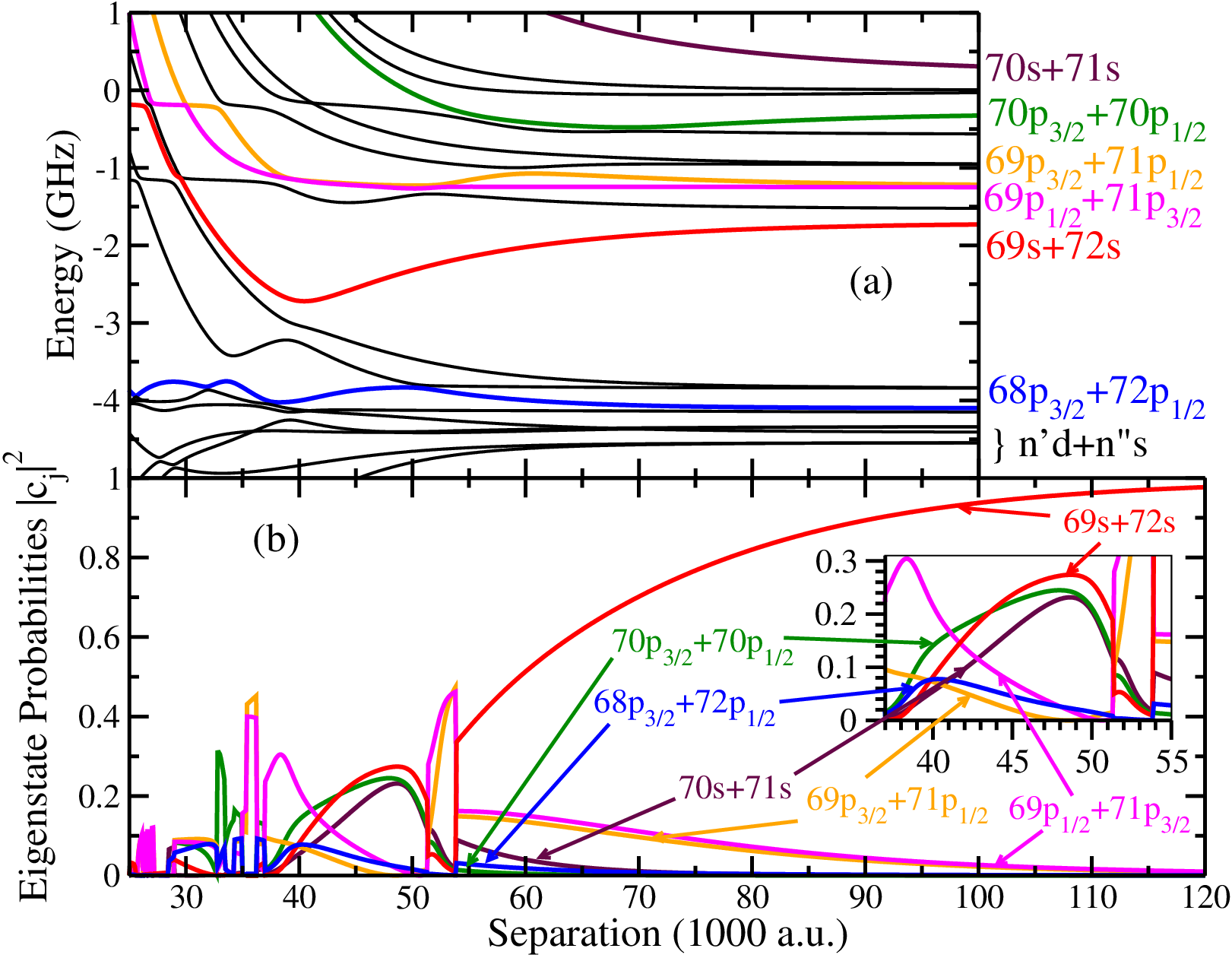}}
\caption{(Color Online) (a) Potential well correlated to the $69s+72s$ asymptote, and 
              (b) its composition, with probabilities $|c_j (R)|^2$ of
              electronic basis $|j\rangle$ vs. $R$ (inset: zoom of the inner region). 
              Same color code used in (a) and (b): see text.}
	   \label{EigVexPlot}
\end{figure}

The molecular electronic states $|\chi_{\lambda}(R)\rangle$ corresponding to the 
potential curves $U_\lambda (R)$ are composed of several molecular electronic 
basis states (\ref{basis}), the exact mixing (after diagonalization) varying with $R$:
\begin{equation}
   |\chi_\lambda (R) \rangle =\sum_j c_j^{(\lambda)}(R) |j\rangle .
   \label{eq:electronic-state}
\end{equation}
Here, $c_j^{(\lambda)}(R)$ are the eigenvectors and 
$|j\rangle$ the electronic basis states. As an example, 
we illustrate the composition of the well correlated to the 
$69s+72s$ asymptote in Fig.~\ref{EigVexPlot}(a).
If we write $|a a'\rangle \equiv |a,a';0_g^+\rangle$, this well is 
composed primarily of $|j \rangle =|69s72s\rangle$, $|70s71s\rangle$, 
$|70p_{3/2}70p_{1/2}\rangle$,  $|69p_{3/2}71p_{1/2}\rangle$, and 
$|69p_{1/2}71p_{3/2}\rangle$; these states are indicated by different 
colors in Fig.~\ref{EigVexPlot}(a). Using the same color labeling, we
show the corresponding probabilities $|c_j (R)|^2$ of all five states against 
$R$ in Fig.~\ref{EigVexPlot}(b). As expected, the right side of the well 
(near the $69s+72s$ asymptote) is composed primarily of the $|69s72s\rangle$ state, 
while the left side contains mostly $|69p_{1/2}71p_{3/2}\rangle$, a state 
whose asymptote is above $69s+72s$. Although one could have 
expected $|68p72p\rangle$ to provide most of the mixing, as suggested by
Fig.~\ref{EigVexPlot}(a), the proximity of the $69s+72s$ and $69p_{1/2}+71p_{3/2}$ 
asymptotes and their strong dipole coupling result in this larger mixing. 
The probability for $|68p_{3/2}72p_{1/2}\rangle$ is
also included in Fig.~\ref{EigVexPlot}(b).

\begin{table}
   \caption{Energies of the deepest bound levels (from the bottom of the well)
                 and classical turning points for the well correlated to the $69s+72s$ 
                 asymptote of the $0_g^+$ symmetry.  
            }
   \begin{tabular}{cccc}
   \hline\hline
   $v$ & Energy (MHz) & $R_1$ (a.u.) & $R_2$ (a.u.) \\
   \hline
    0 & 0.831242 & 40,228 & 40,679  \\
    1 & 2.499032 & 40,068 & 40,849  \\
    2 & 4.166823 & 39,959 & 40,970  \\
    3 & 5.824596 & 39,870 & 41,068  \\
    4 & 7.477361 & 39,795 & 41,154  \\
    5 & 9.125118 & 39,728 & 41,233  \\
   \hline\hline
   \end{tabular}
\label{Lifetimes}
\end{table}   

The coupling to states correlated to lower asymptotes could lead
to predissociation of the macrodimers and their decay into free 
atoms, heating up the sample with more energetic collisions 
and ionization. 
To estimate the predissociation lifetime of our bound states, and because
of the large number of coupled electronic states, we
assume that the total resonance width 
$\Gamma_1$ associated to the nonadiabatic transition from the molecular state 1 is just 
the sum over the widths $\Gamma_{1i}$ associated to transitions to individual adiabatic 
curves $i$, namely $\Gamma_1=\Sigma_i \,\Gamma_{1i}$. To compute the widths,
we use a simple two-channel approach in which the nonadiabatic coupling $V_{1i}$ between 
the close channel corresponding to the molecular bound state (with electronic adiabatic 
basis state $|\chi_1(R)\rangle$), and the open channel corresponding to the dissociation
state (with electronic adiabatic basis state $|\chi_i(R)\rangle$), is 
$V_{1i}(R)=-\frac{m_e}{\mu}  \langle \chi_1 (R)|  \frac{\partial}{\partial R}|\chi_i(R)\rangle  \frac{\partial}{\partial R}$, where $\mu$ is the reduced mass of the system.
We then obtain the width $\Gamma_{1i}$ using a Green's function method \cite{Friedrich}:
$\Gamma_{1i}=2 \pi | \langle \phi_{v}| V_{1i} | \phi_{\rm reg}\rangle |^2$,
where $\phi_{\rm reg}(R)$ is the regular, energy normalized solution of the open channel $i$ 
(in the absence of channel coupling), and the bound vibrational state $\phi_{v}(R)$ 
in the closed channel 1. For the 69$s$72$s$ $0_g^+$ potential curve, this approach give 
predissociation lifetimes that are extremely long, at least $10^{2}$ years, and we thus 
can conclude that these long-range Rydberg molecules are limited by the lifetimes of the 
Rydberg atoms themselves. This result could be expected, since the most important states 
contributing to the well are above the $69s+72s$ asymptote (Fig.~\ref{EigVexPlot}(b)).
We note however that for other wells due to mixing with lower asymptotic electronic electronic 
states, predissociation would play a major role in the lifetime of macrodimers.


%
%

In order to detect and probe the properties of the macrodimers predicted above, one can
photoassociate two ground state atoms using intermediate Rydberg states that provide
good overlap with the target long-range molecular states.  Exciting the two ground state
atoms into an intermediate Rydberg state $ns$, $np$, or $nd$ will allow us to probe 
the different $\ell$-characters (Fig.~\ref{EigVexPlot}) in the bound level.
The photoassociation (PA) rate ${\sf K}_v$ into a bound level $v$ can be calculated
~\cite{Robin98} using
\begin{equation}
   {\sf K}_v \propto I_1 I_2\left|\langle \phi_v|\langle \chi_{\lambda =1}|e^2r_1r_2|
                           \chi_g \rangle |\phi_g\rangle \right|^2 ,
   \label{PARate}
\end{equation} 
where $I_1$ and $I_2$ are the intensities of laser 1 and 2, $|\phi_v(R)\rangle$ 
and $|\chi_{\lambda =1}(R)\rangle$ are the radial and electronic wave functions 
inside the well, respectively, $|\phi_g(R)\rangle$ and $|\chi_g(R) \rangle$ are 
the radial and electronic wave functions of the ground state, respectively, 
and $r_i$ and $e$ are the location and charge of the electron $i$. 
Using expression (\ref{eq:electronic-state}) for $|\chi_{\lambda =1}(R) \rangle$, 
and assuming that $|\chi_g\rangle$ is independent of $R$ 
(corresponding to a flat curve), we can rewrite (\ref{PARate}) as
\begin{equation}
 {\sf K}_v \propto I_1 I_2 \sum_j \left|(d_1d_2)_j\right|^2 \left|\int_0^{\infty}\!\!\!\!
 dR\phi_v^{*}(R)c_j^{*}(R)\phi_g(R)\right|^2 ,
\label{PAInt}
\end{equation} 
where $d_1=\langle n_j\ell_j|er_1|n_g\ell_g\rangle$ and 
$d_2=\langle n'_j\ell'_j|er_2|n'_g\ell'_g\rangle$ are the electronic dipole moments 
between electronic states $|a_g, a'_g\rangle$ and $|a_j,a'_j\rangle$ for atom 1 
and atom 2, respectively.

\begin{figure}[h]
  \centerline{\epsfxsize=3.25in\epsfclipon\epsfbox{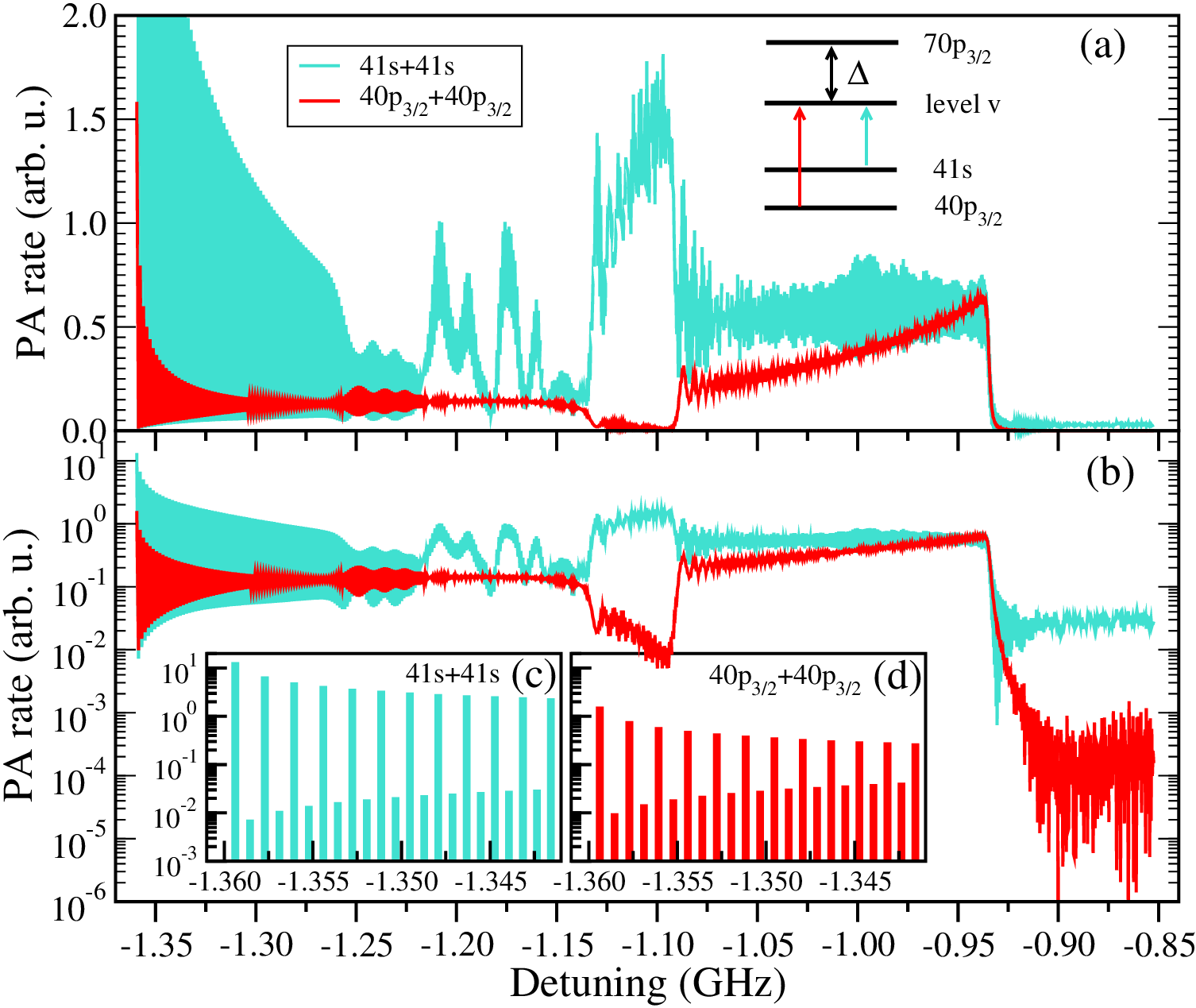}}
\caption{(Color online) PA rate vs. the detuning $\Delta$ from the $70p_{3/2}$ atomic state
              (a) on a linear scale, and (b) on a logarithmic scale. The intermediate state
              $|41s,41s\rangle$ populates the $|np,n'p\rangle$ character in the well (turquoise),
              and the $|40p_{3/2},40p_{3/2}\rangle$ the $|ns,n's\rangle$ (red): (c) and (d) show the
              rate for the deepest levels (see text).  
              }
	   \label{fig:PA-rate}
\end{figure}

We calculate ${\sf K}_v$ by assuming that the Rb $5s$ atoms 
are first excited to an intermediate state that we 
consider to be the electronic ``ground" states $|\chi_g\rangle$, before 
being excited to their final electronic states.  For simplicity, we
consider lower asymptotes in the vicinity of $n=40$ for which 
$|\chi_g\rangle$ does not vary with $R$ (Fig.~\ref{EF00}). 
As shown
in Fig.~\ref{EigVexPlot}, the final electronic state contain mostly $|np,n'p\rangle$ 
or $|ns,n's\rangle$; thus, we assume transitions from $40p_{3/2}+40p_{3/2}$ to
the $ns+n's$ components and from $41s+41s$ to the $np+n'p$ components. 
%
In Fig.~\ref{fig:PA-rate}, we show ${\sf K}_v$ against the detuning $\Delta$ from the atomic
70$p_{3/2}$ levels. For simplicity, we set $\phi_g(R)$ constant, although other choices 
could be made to enhance ${\sf K}_v$, such as a gaussian corresponding to the motion 
of atoms trapped in optical tweezers. The PA rate starting from both atoms in $40p_{3/2}$ is
shown in red, and from $41s$ in turquoise, respectively, on a linear scale in (a) and 
a logarithmic scale in (b). In both cases, the rapid oscillation between a large rate for an
even bound level ($v=0,2,4,\dots$) and a small rate for odd levels 
($v=1,3,5,\dots$) gives the apparent envelop of the signal. This is illustrated
in Figs.~\ref{fig:PA-rate}(c) and (d), where a zoom of the deepest levels $v$ is shown
for both ``ground" states (on a log-scale). This can be understood from the integral in 
Eq.(\ref{PAInt}), which is near zero for an odd wave function $\phi_v(R)$.

The signature of a macrodimer would manifest itself by the appearance of a signal
starting at $\Delta \sim -0.93$ GHz red-detuned from the 70$p_{3/2}$ atomic 
level, and ending abruptly at $\sim -1.36$ GHz. In addition, the PA rate from either 
``ground" state $40p_{3/2}$ or $41s$ can reveal the details of the
$\ell$-mixing in the potential well. First, the overall larger signal for $41s$ 
reflects the scaling of $d_1d_2$ appearing in Eq.(\ref{PAInt}),
with $41s$ being closer than $40p_{3/2}$ to most atomic states appearing in 
the electronic molecular curve (see Fig.~\ref{EigVexPlot}). As $\Delta$ increases
from -1.36 GHz, ${\sf K}_v$ mimic the probabilities $|c_j(R)|^2$. For $41s$, the 
progressive decrease followed by sharp increases between $-1.22$ and $-1.16$ GHz
correspond to the slow decreases of most $|np,n'p\rangle$ components between 
$R\sim 40-50$ a.u. and their sharp increases around $R\sim 33-35$ a.u. (especially from
$|69p_{1/2},71p_{3/2}\rangle$, $|69p_{3/2},71p_{1/2}\rangle$, and
$|70p_{3/2},70p_{1/2}\rangle$; see Fig.~\ref{EigVexPlot}). This is followed by
another large feature between $-1.13$ and $-1.09$ GHz (corresponding to the
sharp increase in $|69p_{1/2},71p_{3/2}\rangle$ and $|69p_{3/2},71p_{1/2}\rangle$
around $R\sim 53,000$ a.u.), and a ``steady" rate before the signal drops to
basically zero. For $40p_{3/2}$, we obtain two obvious features: first, a significant 
drop in ${\sf K}_v$ between $-1.13$ and $-1.09$ GHz which mirrors the decrease
mainly in $|69s,72s\rangle$, and second, a steady growth corresponding to the 
rising probability of $|69s,72s\rangle$ beginning at about $55,000$ a.u. (before the 
signal drops to basically zero).

In conclusion, we have shown the existence of potential wells for the $0_g^+$ symmetry of
doubly-excited atoms due to $\ell$-mixing.  These wells 
support several bound levels separated by a few MHz with lifetimes limited by the 
Rydberg atoms themselves. These vibrational levels could be populated and detected by 
photoassociation. We showed that the signature of macrodimers would be the appearance
of a signal beginning and ending abruptly, and that features in that signal could be
used to probe the $\ell$-character of the potential well. The detection of such 
extended molecules with a lot of internal energy is in itself a goal, and would allow 
to study how {\it macrodimers} are perturbed by surrounding ground state atoms. In 
addition to interesting exotic chemistry ({\it e.g.}, by an approaching Rydberg
atom leading to a strong ultra-long van der Waals complex), the dependence of the PA rate
on the exact $\ell$-mixing (Fig.~\ref{fig:PA-rate}) could potentially be used as a switch 
with application in quantum information ({\it e.g.}, by turning on and off the blockade 
effect \cite{jaksch00}).

The work of N.S. was supported by the National Science Foundation, 
and the work of R.C. in part by the Department of Energy, Office of 
Basic Energy Sciences.  


\end{document}